\documentclass[showpacs,
amssymb, nobibnotes, notitlepage,nofootinbib aps, prd]{revtex4}

\usepackage{amsmath}
\usepackage{amssymb}
\usepackage{amsthm}
\usepackage[pdftex]{color}
\usepackage[pdftex,colorlinks,citecolor=blue,linkcolor=blue,urlcolor=blue]{hyperref} 
\usepackage{graphicx}
\usepackage{dcolumn} 
\usepackage{bm} 
\usepackage{longtable}

\usepackage{ulem}   
\usepackage{comment} 
\usepackage{datetime}

\usepackage{tabularx}
\usepackage{float}
\restylefloat{table}

\newcommand{\beq}{\begin{equation}}
\newcommand{\eeq}{\end{equation}}
\newcommand{\bea}{\begin{eqnarray}}
\newcommand{\eea}{\end{eqnarray}}
\newcommand{\barr}{\begin{array}}
\newcommand{\earr}{\end{array}}

\long\def\begincomment#1\endcomment{}

\begin{document}

%\sffamily

\title{\Large Non-Hermitian two-dimensional harmonic oscillator in noncommutative phase-space
}

%\author{Baloitcha Ezinvi} \email{ezinvi.baloitcha@cipma.uac.bj}  
%\affiliation{International Chair in Mathematical Physics and Applications (ICMPA-UNESCO Chair), University of Abomey-Calavi,
%072B.P.50, Cotonou, Republic of Benin}

\author{Emanonfi Elias N'Dolo}
\email{emanonfieliasndolo@yahoo.fr}
\affiliation{International Chair in Mathematical Physics and Applications (ICMPA-UNESCO Chair), University of Abomey-Calavi,
072B.P.50, Cotonou, Republic of Benin}

\date{\today}

\begin{abstract}
In this paper, we extend the result of [{\bf Andreas Fring et al} J. Phys. A \textbf{43}, 345401 (2010)] in noncommutative phase-space (NCPS). We compute the non-Hermitian Hamiltonian of a harmonic oscillator in NCPS. We construct a new $PT$-symmetry in noncommutative phase-space and prove that the system does not possess a broken $PT$-regime. We then compute the eigenvalue spectrum of the non-Hermitian Hamiltonian of the system. %We construct the Galilei unitary group with this non-Hermitian variable. 


\end{abstract}
\maketitle
\section{Introduction}
%{\color{red}
The modern physics exists to put together the physics of small mass (quantum mechanics) and the physics of large mass (general relativity), i.e the study of physics near the Planck length $\lambda_p = \sqrt{\frac{G\hbar}{c^3}} \approx 1, 6. 10^{-35} m$. In the last century, several approaches have been given to tackle this unification. Among them, there exist very promising theories such as string theory,  noncommutative geometry (NCG), loop quantum gravity, and tensorial group field theory.

Noncommutative geometry started with the work of Snyder \cite{Snyder:1946qz} in quantum mechanics and has been more rigorously defined by Connes \cite{Connes:1994yd}. In string theory, it is considered as the static solution of string dynamics \cite{Seiberg:1999vs}. NCG arises as a possible scenario for the short-distance of physical theories in the Planck length (see \cite{Doplicher:1994tu} and references therein).
As in ordinary quantum mechanics, the NC coordinates satisfy the coordinate-coordinate version of the Heisenberg uncertainty relation, namely $\Delta x^\mu\Delta x^\nu\geq \theta$, and then makes the spacetime a quantum space. This idea leads to the concept of quantum gravity, since quantizing spacetime leads to quantizing gravity.

With this new domain of quantum gravity, many problems in physics give openness to researchers. For instance, the {\it Lorentz violation} in spacetime more than two dimensions \cite{Ferrari:2005ng}, which, in part, modifies the dispersion relations \cite{Jackiw:2001dj}, the Landau problem reveals when a quantum oscillator interacts with a magnetic field.
The noncommutative phase-space has been studied in \cite{BenGeloun:2009hkc}-\cite{Gnatenko:2019qln} to restore the time-reversal and rotational symmetries.  The noncommutative momenta problem is often interpreted as the well-known Landau problem in the presence of a harmonic interaction in the noncommutative Moyal plane. This is because the presence of noncommutative momenta mimics the presence of a magnetic field in the system since the gauge covariant derivatives can be interpreted as the noncommutative momenta in this setting. The phase space structure is also thought to be a more consistent approach for unifying quantum mechanics and general relativity high energy scales \cite{Townsend:1977xw, Majid:1988we}.

Quantum oscillator in two-dimensions is still revealing many interesting problems due to the degeneracy appearing in the spectrum of the eigenvalue. In particular, two-dimensional noncommutative harmonic oscillators have attracted a great deal of attention in the literature \cite{Nair:2000ii}-\cite{Giri:2008qu}. It is shown in \cite{BenGeloun:2009hkc} that the time-reversal symmetry of the two-dimensional harmonic oscillator Hamiltonian gets stored since the energy spectrum becomes degenerate.  Furthermore, it is shown in \cite{Bose:1994sj, Horvathy:2010wv} that a nonrelativistic system in $(2+1)$-dimensions admitting fractional spin can exhibit Galilean symmetry through a two-fold central extended Galilean algebra, where one involves the commutator of the boost generators $K_i$ between themselves, which is a non-vanishing constant $[K_1, K_2] = i\hbar\kappa$, and the others involve boost and linear momentum, which gives, the mass $m$: $[K_i, P_j] =i\hbar m\delta_{ij}$, with other commutators taking their usual forms.

In quantum mechanics, the non-Hermitian operator has been interpreted for a long time as no sense physically objects up to the pioneering work of Bender and Boettcher \cite{Bender:1998ke}. The authors showed that there exist the non-Hermitian operators whose spectra are positive and real. Moreover, it is shown in \cite{Mostafazadeh:2001jk, Bagchi:2009wb, Bender:2007nj} that the Hermitian condition can be replaced by the weaker condition parity and time reversing $PT$-symmetry. It is proven in \cite{Bagchi:2009wb} that non-Hermitian Hamiltonian and the deformed quantum mechanics based on deformations of the uncertainty relation may be treated in a similar fashion. The authors demonstrated that the deformed Heisenberg algebra leads to non-Hermitian Hamiltonian, different minimal uncertainties, minimal lengths, and minimal momenta.  In noncommutative frame, the large number of possible free parameters in the calculation of deformed oscillators algebra using the canonical method is reduced to a manageable amount by imposing various different versions of $PT$-symmetry on the underlying spaces, which are dictated by the specific physical problem \cite{Dey:2012dm}. In \cite{Giri:2008iq}, it has been showed that the $PT$- symmetry is only possible for complex noncommutative parameters. Contrary to this work, the authors in \cite{Fring:2010pw,Dey:2012dm} have proven that it is indeed possible to implement $PT$-symmetry on noncommutative spaces while keeping the noncommutative parameters real. Furthermore, the stringent restriction of the preserved rotational and time-reversal symmetries on the minimal momentum in phase space is found recently in \cite{Gnatenko:2018gqa}.

The position-dependent noncommutativity is used in \cite{Fring:2010pw} to investigate about non-Hermitian Hamiltonian. The authors used only the position-position noncommutative algebra to prove that the variable associated is non-Hermitian. They constructed the $P_\theta T-$symmetry which leaves invariant the Hamiltonian of a system.

. % about noncommutative position and momentum deformation  

The main aim of this paper is to extend the work in \cite{Fring:2010pw} into the deformation of the momentum-momentum noncommutative constant. 
This work is organized as follows. In Section \eqref{pts}, the $PT-$symmetry in noncommutative phase-space is constructed. In Section \eqref{nho}, the position-dependant deformation of noncommutative Lie algebra is given and the minimal length and the large class of this minimal length which correspond to the {\it squeezing} state are constructed. In Section \eqref{PT}, the Hamiltonian of the harmonic oscillator in these new variables of the function depending on hermitian variables is computed.  We factorized this Hamiltonian in the function of new scale operators and its eigenvalue is calculated. The work is ended by concluding remarks. 
%n physics, the symmetry is an important objects which is used to simplify the computation.

%The noncommutatif theory is one of the modern physics which take care of the study of physics near the Planck length.

%}
\section{PT-symmetry in noncommutative phase-space}\label{pts}
Consider the noncommutative phase-space algebra as 
\bea\label{ca}
[x, y] &=& i\theta;\qquad [x, p_x] = i\hbar;\qquad [y, p_y] = i\hbar;\cr [p_x, p_y] &=& i\eta;\qquad [x, p_y] = 0;\qquad [y, p_x] = 0
\eea
where $(\theta, \eta) \in \mathbb R^2$ are the noncommutative parameters. Using the Bopp-shift in deformation phase-space \cite{Zhang:2004yu,Hounkonnou:2011gx}
\bea\label{bs}
\hat x_i = q_i -\frac{\theta}{2\hbar}\epsilon_{ij}\pi_j;\qquad \hat p_i = \pi_i +\frac{\eta}{2\hbar}\epsilon_{ij} q_j; \quad \epsilon_{ij} = -\epsilon_{ji} =1;\quad i, j = 1, 2,
\eea
where the standard coordinates $q_i, \pi_i$ are commutative coordinates satisfying $[q_i, \pi_j] = \delta_{ij}$, we can then construct the $PT$-symmetry in noncommutative phase-space. The $PT$-symmetry is the simultaneous parity $P$ and time $T$ reversal transformations. In ordinary space, $Pq_i \mapsto-q_i$;\quad $P\pi_i \mapsto -\pi_i$,\quad $Tq_i \mapsto q_i$;\quad $T\pi_i\mapsto -\pi_i$. Contrary to the work in \cite{Giri:2008iq}, the authors in \cite{Fring:2010pw,Dey:2012dm} have proven that it is indeed
possible to implement $PT$-symmetry on noncommutative spaces while keeping the noncommutative constants real. They found that $P_\theta T$ can be construted if the  noncommutative parameter is transformed as $\theta\mapsto -\theta$.  We extend this work in noncommutative phase-space. The $P_{\theta\eta}T$ is possible if $\eta \mapsto -\eta$. In order to have a total reflection of the positions (i.e. in x-direction and y-direction), we obtain the $PT$-symmetry as follows:
\bea\label{pt}
&&PT: q_j \mapsto -q_j, \quad \pi_j\mapsto \pi_j,\quad i\mapsto -i, \quad j = 1, 2.\cr
&&P_\theta T: \hat x_j\mapsto -\hat x_j, \quad \hat p_j \mapsto \hat p_j,\quad \theta \mapsto -\theta \quad i\mapsto -i\qquad j = 1, 2.\cr
&&P_{\theta\eta}T:\hat x_j\mapsto -\hat x_j, \quad \hat p_j \mapsto \hat p_j,\quad \theta \mapsto -\theta,\quad \eta\mapsto  -\eta,\quad i\mapsto -i\qquad j = 1, 2. \label{pts}
\eea
Let's stress that the third line of the transformation \eqref{pt} leaves invariant the noncommutative relations \eqref{ca}.
\section{Non-Hermitian operators in noncommutative phase-space}\label{nho}
Consider the two-dimensional non-Hermitian operators in noncommutative phase-space $(X, Y, P_x , P_y)$ where the constants parameters $\theta$ and $\eta$ currently used, are replaced by the position-dependent parameters $\theta(X,Y) = \theta(\mathbb I+\tau Y^2)$ and $\eta(X,Y) =\eta(\mathbb I+\tau Y^2)$ \cite{Fring:2010pw}. These operators satisfy the following relations
\bea
&&[X, Y] = i\theta(\mathbb I +\tau Y^2);\quad [X, P_x] = [Y, P_y] = i\hbar(\mathbb I+\tau Y^2); \cr &&[X, P_y] = 2i\tau Y (\theta P_y+\hbar X);\quad [P_x, P_y] = i\eta(\mathbb I+\tau Y^2); \quad [Y, P_x] =0
\eea
where $\theta, \eta$ are the noncommutative phase-space parameters and $\tau$ is the non-Hermitian parameter. Let us recall that, the variables in capital letters are for non-Hermitian and the variables in small letters are for noncommutative phase space.
In the limit $\tau\longrightarrow 0$, we recover the flat noncommutative phase-space algebra (1). To ensure consistency of this deformation algebra, we emphasize that the Jacobi identities are satisfied :
\bea
[A, [B, C]] + [B, [C, A]] +[C, [A, B]] = 0  \quad \mbox {for}\quad A, B, C \in\{X, Y, P_x, P_y\}.
\eea
However, we can consider the linear function of $\theta(X,Y) =\theta (\mathbb I+\tau Y)$, $\eta(X,Y) = \eta(\mathbb I+\tau Y)$, with $\theta, \eta,\tau<<1$, and can define the commutation relations to satisfy the Jacobi identities but it is not possible to obtain the minimal length in this case.\\
Like a Bopp-Shift transformation, we can express  the non-Hermitian variables in terms of the Hermitian variables :
\bea
X = (1+\tau y^2)x, \quad Y = y,\quad P_x = p_x,\quad P_y = (1+\tau y^2)p_y. %+i\gamma p_x.
\eea

To construct the adjoint of above operators, we must use the Dyson map $\zeta = (1+\tau Y^2)^{-1/2}$.
The adjoint of these operators are given by $\mathcal O^\dagger = \zeta \mathcal O\zeta^{-1}$. We obtain the following operators 
%{\color{blue

\bea\label{inp}
X^\dagger = X+2i\tau \theta Y,\quad Y^\dagger = Y; \quad P_x^\dagger  = P_x; \quad P_y^{\dagger} = P_y-2i\tau \hbar Y. %+ i\gamma P_x.
\eea
%{\color{blue}
	In general, these operators are not Hermitian. To recover the Hermitian inner product for the restored Hilbert space, let's define like in \cite{Fring:2010pw}, a new metric operator $\rho = \zeta^\dagger\zeta$ and inner product $\langle .\vert .\rangle_\rho$. Then, the inner product between the arbitrary function $\vert \Phi\rangle$ and $\vert \Psi\rangle$ can be defined as:
	\bea\label{INP}
	\langle \Phi\vert \psi\rangle_\rho :=\langle\Phi\vert\rho\Psi\rangle = \int^{+\infty}_{-\infty}\frac{dy}{1+\tau y^2}\psi^{*}(y)\phi(y).
	\eea
	Then, any observable operator $\mathcal O$ has to satisfy
	\bea
	\langle \Phi \vert \mathcal O\Psi\rangle_\rho = \langle \mathcal O\Phi\vert \Psi\rangle_\rho,
	\eea
which makes the operator $\mathcal O$ Hermitian in this Hilbert space.

The identity operator can thus be expanded as
\bea\label{Id}
 \int^{+\infty}_{-\infty}\frac{dy}{1+\tau y^2}\vert y\rangle\langle y\vert = \mathbb I.
\eea
 The scalar product between two positions variables is therefore writing
\bea\label{sp}
\langle y\vert y'\rangle = (\mathbb I+\tau y^2)\delta(y-y').
\eea
Let's stress that the relations \eqref{Id} and \eqref{sp} are analog to that we use in ordinary quantum mechanics, which is recovered for $\tau \longrightarrow 0$.
%}
\section{General class of minimum-uncertainty in noncommutative phase-space}

	Consider two operators $A$ and $B$. The uncertainty relation is given by
	\bea
	\Delta A \Delta B\geq \frac{1}{2}\vert \langle [A, B]\rangle_\rho\vert.
	\eea 
	Then, by applying this identity to the non-Hermitians position operators \eqref{inp}, under the inner product \eqref{INP}, we get
	\bea
	\Delta X\Delta Y \geq \frac{1}{2}\vert i\theta\langle \mathbb I +\tau Y^2\rangle \vert.
	\eea

To obtain the minimum length, let's define the function 
\bea\label{f}
f(\Delta X, \Delta Y) = \Delta X \Delta Y -\frac{\theta}{2}(\mathbb I +\tau \langle Y\rangle^2+\tau(\Delta Y)^2).
\eea
Starting with a simultaneous $X, Y$-measurement and using the standard method argument for minimizing, we have to solve the equation:
	\bea
	\partial_{\Delta Y}f(\Delta X, \Delta Y) =0\qquad \mbox{and}\qquad f(\Delta X,\Delta Y) =0.
	\eea

We obtain from the expression \eqref{f} that
\bea
\Delta X_{min} =\theta\sqrt \tau \sqrt{1+\tau\langle Y\rangle^2}.
\eea
In the same way, we can compute the minimal momentum to measure simultaneous $Y$ and $P_y$ as follows
\bea
(\Delta P_y)_{min} = \hbar\sqrt\tau\sqrt{1+\tau\langle Y\rangle^2}.
\eea
 The most solutions of simultaneous measurements of $Y, P_y$ are given by 
%{\color{red}
\bea
\Delta Y =\frac{\Delta P_y}{\hbar\tau}\pm\frac{\sqrt{\Delta P_y^2-\hbar^2\tau(1+\tau \langle Y\rangle^2)}}{\hbar\tau}.
\eea
The general class of minimum-uncertainty corresponding to the so-called {\it squeezing state} of measure $Y, P_y$ is obtained by setting
\bea
(\Delta Y)^2<\vert\langle [Y,P_y]\rangle_\rho\vert.
\eea
Then, the general class of states which may have reduced uncertainty in one quadrature at the expense of increased uncertainty is given by
\bea
\Delta Y
\in \Bigg[0, \sqrt{\frac{\hbar}{2-\hbar\tau}(1+\tau\langle Y\rangle^2)}\Bigg[.
\eea
For vanishing of the non-hermitian parameter $\tau\longrightarrow 0$, we obtain the ordinary minimal uncertainty general class for the existing squeezing state of $Y, P_y$-simultaneous measure.
%For two functions $\psi$ and $\phi$, the scalar product is such that:

\section{$P_{\theta\eta}T$-symmetry with Non-Hermitian Hamiltonian}\label{PT}
In section \eqref{nho}, we have given the $P_{\theta\eta}T$-symmetry in noncommutative phase-space. The natural question that arises is: what are the properties of Hamiltonian systems on this space? In this section, we will answer this question and then highlight the importance of non-Hermitian theory by computing the eigenvalue of the non-Hermitian Hamiltonian
\subsection{Harmonic oscillator in noncomutative phase-space}
Consider the harmonic oscillator described by the non-Hermitian phase-space variables 
\bea\label{hnh}
H = \frac{1}{2m}(P_x^2+P_y^2)+\frac{1}{2}m\omega^2(X^2+Y^2).
\eea
To obtain the Hermitian Hamiltonian, let us first write it in terms of noncommutative variables 
%{\color{red}
\bea\label{He}
%H &=& \frac{1}{2m}(p_x^2+p_y^2)+ \frac{1}{2} m\omega^2 (x^2+y^2) +(2\tau y^2 +\tau^2 y^4)\big(\frac{1}{2m}p_y^2 +\frac{1}{2} m\omega^2 x^2\big)\\
H &=& \frac{1}{2m}\Big[p_x^2+(1+\tau y^2)^2 p_y^2 -2i\hbar \tau y(1+\tau y^2)p_y\Big]+\frac{1}{2}m\omega^2 \Big[(1+\tau y^2)^2 x^2+2i\tau\theta y(1+\tau y^2) x+y^2\Big]\cr
&=&\frac{1}{2m}\Big[p_x^2+(1+2\tau y^2)p_y^2-2i\hbar\tau yp_y\Big]+\frac{1}{2}m\omega^2\Big[(1+2\tau y^2) x^2 +2i\tau \theta y x +y^2\Big] + \mathcal O(\tau^2).%+\mathcal O(\tau^2)
\eea

%}
Then, it should be noted that the non-Hermitian system can be regarded as a sum of the Hermitian system along with the entanglement of the interaction system, which arises from the Hermitian parameter $\tau$.  

To find the eigenvalue of the system, let us rewrite \eqref{He} in terms of the ordinary variables. Thus, using the Bopp-Shift transformation \eqref{bs}, %recalled in recently work \cite{Biswas:2020obt} :
%\color{red}
%\bea
% x_i = q_i-\frac{\theta}{2\hbar}\epsilon_{ij}\pi_j 
%;\quad p_i = \pi_i +\frac{\eta}{2\hbar}\epsilon_{ij} q_j,
%\eea
the Hamiltonian \eqref{He} in ordinary phase-space can be writte as follows :
\bea\label{H}
H
&=& \frac{1}{2m_R}\big(\pi_1^2+\pi_2^2\big)+\frac{1}{2}m\omega^2_R\big( q_1^2 +q_2^2\big) +\Big(\frac{\eta}{2m\hbar}+\frac{m\omega\theta}{2\hbar}\Big)(q_2\pi_1-q_1\pi_2\big)
\cr &&+ \tau\Bigg[\Big(m\omega^2 +\frac{\eta^2}{4m\hbar^2}\Big)q_2^2q_1^2+\Big(\frac{1}{m}+\frac{m\omega^2\theta^2}{4\hbar^2}\Big)q_2^2\pi_2^2-\Big(\frac{\eta}{m\hbar}-\frac{m\omega^2\theta^2}{\hbar^2}\Big)q_2^2q_1\pi_2\cr&& +\Big(\frac{m\omega^2\theta}{\hbar}+\frac{\theta\eta^2}{4m\hbar^3}\Big)\big(2i\hbar q_2q_1+q_2q_1^2\pi_1\big)+\Big(\frac{m\omega^2\theta^2}{4\hbar^2}+\frac{\theta^2\eta^2}{16m\hbar^2}\Big)\big(-2\hbar+4i\hbar q_1\pi_1+q_1^2\pi_1^2\big)\cr&&+\Big(\frac{\theta\eta}{m\hbar}-\frac{m\omega^2\theta^2}{\hbar^2}\Big)\big(i\hbar q_2\pi_2+q_2q_1\pi_1\pi_2\big)-\Big(\frac{i\hbar}{m}+\frac{m\omega^2\theta }{2\hbar}\Big)q_2\pi_2+\Big(\frac{i\eta}{2m}+im\omega^2\theta\Big)q_2q_1\cr&&+\Big(\frac{i\theta\eta}{4m\hbar}+\frac{im\omega^2\theta^2}{2\hbar}\Big)\big(i\hbar+q_1\pi_1\big)+\frac{i\theta }{2m}\pi_1\pi_2+\frac{\theta}{m\hbar}q_2\pi_1\pi_2^2+\frac{\theta^2}{4m\hbar^2}\pi_1^2\pi_2^2\cr&&-\frac{\theta\eta^2 }{4m\hbar^3}\big(2i\hbar\pi_1\pi_2 +q_1\pi_1^2\pi_2\big)\Bigg] +\mathcal O(\tau^2),
\eea 
where
\bea
\frac{1}{m_R} = \frac{1}{m}+\frac{m\omega^2\theta^2}{2\hbar} = \frac{\kappa}{m};\qquad\kappa = \sqrt{1+\frac{m^2\omega^2\theta^2}{4\hbar^2}};\quad \omega_R = \omega\sqrt{1+\frac{\eta^2}{4m^2\omega^2\hbar^2}}.
\eea
In the limit $\tau =0$, we obtain the Hermitian noncommutative phase-space Hamiltonian, and recover the result of \cite{Biswas:2020obt}. Furthermore, if $\eta =0$, we get the result obtained in \cite{Fring:2010pw}. Let us note that, the Hamiltonian $\eqref{H}$ is $P_{\theta\eta}T$ symmetric.
%%%%%%%%%%%%%%%%%%%%%%%%%%%%%%%%%%%%%%%%%%%%
\subsection{Eigenvalue of non-Hermitian operators of two-dimensional harmonic oscillator}
%%%%%%%%%%%%%%%%%%%%%%%%%%%%%%%%%%%%%%%%%%%%%%%%%%%
 At this step, since $\theta, \eta, \tau$ are chosen very small, we can simplify our Hamiltonian \eqref{H} in the form :
\bea\label{Hap}
H = H_c + H_{\theta,\eta} + H_{\tau} + \mathcal O(\tau\eta, \tau \theta, \theta\eta, \tau^2, \eta^2, \theta^2)
\eea
with 
\bea\label{sha}
\left\{\begin{array}{ccc} 
H_c = \frac{1}{2m}\big(\pi_1^2+\pi_2^2\big)+\frac{1}{2}m\omega^2\big( q_1^2 +q_2^2\big)\cr H_{\theta,\eta} = \Big(\frac{\eta}{2m\hbar}+\frac{m\omega^2\theta}{2\hbar}\Big)(q_2\pi_1-q_1\pi_2\big)\cr H_{\tau}=\tau(m\omega^2 q_2^2q_1^2-\frac{i\hbar}{m}q_2 \pi_2+\frac{1}{m}q_2^2\pi_2^2).\end{array}\right.
\eea
Let's remark that $[H_c, H_{\theta,\eta}] = 0$ and $[H_\tau, H_{\theta,\eta}] =0$. Then, the Hamiltonian $H_{\theta,\eta}$ is a generator of dynamical symmetry group $SO(2)$ corresponding to our Hamiltonian system and can help to diagonalize simultaneously these operators. 
To determine the corresponding basis, let's consider the new following Fock algebra generators : 
\bea\label{sco}
\left\{\begin{array}{cccc}
A_+ = \frac{1}{2}\sqrt{\frac{m\omega}{\hbar}}[q_1+iq_2+\frac{i}{m\omega}(\pi_1+i\pi_2)]\cr
A_{-} = \frac{1}{2}\sqrt{\frac{m\omega}{\hbar}}[q_1-iq_2+\frac{i}{m\omega}(\pi_1-i\pi_2)]\cr
A_{+}^{\dagger} = \frac{1}{2}\sqrt{\frac{m\omega}{\hbar}}[q_1-iq_2-\frac{i}{m\omega}(\pi_1-i\pi_2)]\\
A_{-}^{\dagger} = \frac{1}{2}\sqrt{\frac{m\omega}{\hbar}}[q_1+iq_2-\frac{i}{m\omega}(\pi_1+i\pi_2)].\end{array}\right.
\eea
The commutations relations of these scale operators are given by
\bea
[A_+, A_{+}^{\dagger} ] = \mathbb I = [A_-, A_{-}^{\dagger}] ;\quad \mbox{and all the others vanish}.
\eea
The helicity Fock basis coreesponding is $\big\{\vert n_+, n_-\rangle\big\}$ such that :
\bea
A_+\vert n_+, n_-\rangle &=& \sqrt{n_+}\vert n_+-1, n_-\rangle,\quad A_+^\dagger\vert n_+, n_-\rangle = \sqrt{n_++1}\vert n_+ +1, n_-\rangle,\cr A_-\vert n_+, n_-\rangle &=& \sqrt{n_-}\vert n_+, n_- -1\rangle, \quad A_-^\dagger\vert n_+, n_-\rangle = \sqrt{n_- +1}\vert n_+, n_- +1\rangle.
\eea
The eigenfunction $\vert\psi_{n_+, n_-}\rangle$ is related to the background basis $\vert\psi_{0, 0}\rangle$ by :
\bea
\vert \psi_{n_+, n_-}\rangle = \frac{1}{\sqrt{n_+! n_-!}}\Big(A_+^{\dagger}\Big)^{n_+}\Big(A_-^{\dagger}\Big)^{n_-}\vert \psi_{0, 0}\rangle.
\eea
Also the eigenfunctions are $PT$-symmetric and hence there is no broken $PT$-regime.
In this basis, the eigenvalue of each part of the system of Hamiltonian \eqref{sha} is given by
\bea
 E_c^{n_+, n_-} = \hbar\omega(n_+ + n_- +1),\quad E_{\theta, \eta}^{n_+, n_-} = \hbar\Big(\frac{\eta}{2m}+\frac{m\omega^2\theta}{2}\Big)(n_+ - n_-), \quad E_{\tau}^{n_+, n_-} = \frac{\tau\hbar^2}{2m}\Big(2n_+ n_- +n_+ +n_- +2\Big), \nonumber\\ n_+> n_-.
\eea

Then, the energy spectrum of the non-Hermitian operators of two-dimensional oscillator in noncommutative phase space is given by :
\bea
E_{\theta, \eta, \tau}^{n_+, n_-} = E_c^{n_+, n_-} + E_{\theta,\eta}^{n_+, n_-}+E_\tau^{n_+, n_-}, \qquad n_+, n_- \in \mathbb N.
\eea

First of all, we remark that this energy is real, and then answer the famous question of Bender \cite{Bender:1998ke}. In addition, if the non-Hermitian parameter vanishes $\tau\longrightarrow 0$, we obtain the same result as found in the non-commutative phase space with the Hermitian operators \cite{Muthukumar:2002cn, Scholtz:2008zu, BenGeloun:2009hkc, Lawson:2020vbd}.
%%%%%%%%%%%%%%%%%%%%%%%%%%%%%%%%%%%%%%%%%%%%%%%%%%%%%%%%
%\subsection{Factorization of non-Hermitian Hamiltonian in two-dimensional harmonic oscillator}

%%%%%%%%%%%%%%%%%%%%%%%%%%%%%%%ù%%%%%%%%%%
%{\color{blue}

\section{Concluding remarks}
In this paper, the non-Hermitian two-dimensional harmonic oscillator in noncommutative phase space has been extended. The parity-time reversal symmetry in noncommutative phase space, which preserves the Hamiltonian of the system, has been constructed. The energy spectrum of this kind of the harmonic oscillator has been computed and found to be real, confirming the theory of Bender \cite{Bender:1998ke}. In the limit of the non-Hermitian parameter $\tau =0$, we obtain the result of the noncommutative phase space with $PT$-symmetry in the previous investigations. Furthermore, in the case where the parameter $\eta= 0$, we obtain the spectrum of the harmonic oscillator in noncommutative space like in \cite{Scholtz:2008zu} and for all vanish parameters, we obtain the ordinary eigenvalue of a harmonic oscillator. \\ 
For the next work, we plan to construct the {\it squeezed states} corresponding to the general class of minimum-uncertainty.

\section*{Acknowledgments}
I would like to thank Professor Andreas Fring for fruitful discussions. I am grateful.
\section*{Appendix}
Here we give some lines of eigenvalue calculation. 

Conversery from \eqref{sco}, we obtain :
\bea\label{pm}
\left\{\begin{array}{cccc}
q_1 = \sqrt{\frac{\hbar}{4m\omega}}\big(A_+ + A_- + A_{+}^{\dagger}+A_-^{\dagger}\big)\cr
q_2 = \frac{1}{2i}\sqrt{\frac{\hbar}{m\omega}}\big(A_+ - A_- -A_+^{\dagger}+A_{-}^\dagger\big)\cr
\pi_1 = \frac{\sqrt{\hbar m\omega}}{2i}\big(A_+ + A_- -A_+^{\dagger}-A_{-}^{\dagger}\big)\cr
\pi_2 = -\frac{\sqrt{\hbar m\omega}}{2}\big(A_+ - A_- + A_+^{\dagger} -A_-^{\dagger}).
\end{array}\right.
\eea
 We have: 
\bea
q_1^2 &=& \frac{\hbar}{4m\omega}\Big[A_+^2 +A_+A_- +A_+A_+^\dagger +A_+A_-^\dagger+A_-A_+ +A_-^2+A_-A_+^\dagger+A_-A_-^\dagger +A_+^\dagger A_+ + A_+^\dagger A_- +( A_+^\dagger)^2 +A_+^\dagger A_-^\dagger \cr&&+A_-^\dagger A_+ + A_-^\dagger A_- + A_-^\dagger A_+^\dagger +(A_-^\dagger)^2\Big]\\
q_2^2 &=& -\frac{\hbar}{4m\omega}\Big[A_+^2 - A_+A_- - A_+ A_+^\dagger + A_+ A_-^\dagger - A_-A_+ + (A_-)^2 + A_-A_+^\dagger -A_-A_-^\dagger-A_+^\dagger A_+ + A_+^\dagger A_- + (A_+^\dagger)^2 -A_+^\dagger A_-^\dagger \cr&&+ A_-^\dagger A_+ -A_-^\dagger A_- -A_-^\dagger A_+^\dagger + (A_-^\dagger)^2\Big]\\
\pi_1^2 &=&-\frac{\hbar m\omega}{4}\Big[ A_+^2 + A_+A_- -A_+A_+^\dagger -A_+A_-^\dagger+A_-A_++A_-^2 - A_-A_+^\dagger -A_-A_-^\dagger -A_+^\dagger A_+ -A_+^\dagger A_- + (A_+^\dagger)^2+A_+^\dagger A_-^\dagger \cr&&- A_-^\dagger A_+ -A_-^\dagger A_- + A_-^\dagger A_+^\dagger +(A_-^\dagger)^2\Big]\\
\pi_2^2 &=& \frac{\hbar m\omega}{4}\Big[A_+^2 -A_+A_- + A_+ A_+^\dagger -A_+A_-^\dagger-A_-A_+ +A_-^2 -A_-A_+^\dagger +A_-A_-^\dagger+A_+^\dagger A_+ -A_+^\dagger A_- +(A_+^\dagger)^2-A_+^\dagger A_-^\dagger \cr &&-A_-^\dagger A_++A_-^\dagger A_- - A_-^\dagger A_+^\dagger +(A_-^\dagger)^2\Big]
\eea 
In Fock basis, we have
\bea
m\omega^2q_1^2 q_2^2\vert n_+, n_-\rangle &=& \frac{\hbar^2}{4m}\Big[2n_+ n_- + n_+ +n_- +\frac{3}{2}\Big]\vert n_+, n_-\rangle,\qquad -\frac{i\hbar}{m}q_2\pi_2\vert n_+, n_-\rangle = \frac{\hbar^2}{2m}\vert n_+, n_-\rangle\cr
\frac{1}{m}q_2^2\pi_2^2 \vert n_+, n_-\rangle &=& \frac{\hbar^2}{4m}\Big[2n_+n_- +n_+ + n_- +\frac{1}{2}\Big]\vert n_+, n_-\rangle
\eea
Then
\bea
H_\tau\vert n_+, n_-\rangle &=& \tau\Big(m\omega^2 q_2^2q_1^2-\frac{i\hbar}{m}q_2 \pi_2+\frac{1}{m}q_2^2\pi_2^2\Big)\vert n_+, n_-\rangle=\frac{\tau\hbar^2}{2m}\Big(2n_+n_- + n_+ + n_- +2\Big)\vert n_+, n_-\rangle \\
H_{c}\vert n_+, n_-\rangle &=& \hbar\omega(n_++n_-+1)\vert n_+, n_-\rangle, \quad H_{\theta,\eta}\vert n_+, n-\rangle = \hbar\Big(\frac{\eta}{2m}+\frac{m\omega^2\theta}{2}\Big)(n_+ - n_-)\vert n_+, n_-\rangle.
\eea

\end{document}